\newcommand{\CLASSINPUTtoptextmargin}{0.75in}
\newcommand{\CLASSINPUTbottomtextmargin}{1.01in}
\documentclass[conference]{IEEEtran}
\IEEEoverridecommandlockouts
\usepackage{cite}
\usepackage{amsmath,amssymb,amsfonts,gensymb}
\usepackage{algorithmic}
\usepackage{graphicx}
\usepackage{textcomp}
\usepackage{xcolor}
\usepackage{hyperref}
\usepackage{booktabs}
\usepackage{multirow}
\usepackage{adjustbox}
\usepackage[absolute,showboxes]{textpos}
\TPMargin{5pt}

\newcommand{\copyrightstatement}{
    \begin{textblock}{0.84}(0.08,0.93)    
         \noindent
         \footnotesize
         \copyright  2022 IEEE.  Personal use of this material is permitted.  Permission from IEEE must be obtained for all other uses, in any current or future media, including reprinting/republishing this material for advertising or promotional purposes, creating new collective works, for resale or redistribution to servers or lists, or reuse of any copyrighted component of this work in other works. The peer-reviewed paper is available at: \url{https://doi.org/10.1109/ICECTA57148.2022.9990500}
    \end{textblock}
}

\begin{document}
\copyrightstatement
\bstctlcite{IEEEexample:BSTcontrol}
\title{Martian Ionosphere Electron Density\\Prediction Using Bagged Trees}

\author{\IEEEauthorblockN{Abdollah Masoud Darya, Noora Alameri, Muhammad Mubasshir Shaikh and Ilias Fernini}
\IEEEauthorblockA{SAASST,
University of Sharjah\\
Sharjah, United Arab Emirates\\
adarya@sharjah.ac.ae, nalameri@sharjah.ac.ae, mshaikh@sharjah.ac.ae, ifernini@sharjah.ac.ae}}

\maketitle

\begin{abstract}
The availability of Martian atmospheric data provided by several Martian missions broadened the opportunity to investigate and study the conditions of the Martian ionosphere. As such, ionospheric models play a crucial part in improving our understanding of ionospheric behavior in response to different spatial, temporal, and space weather conditions. This work represents an initial attempt to construct an electron density prediction model of the Martian ionosphere using machine learning. The model targets the ionosphere at solar zenith ranging from 70 to 90 degrees, and as such only utilizes observations from the Mars Global Surveyor mission. The performance of different machine learning methods was compared in terms of root mean square error, coefficient of determination, and mean absolute error. The bagged regression trees method performed best out of all the evaluated methods. Furthermore, the optimized bagged regression trees model outperformed other Martian ionosphere models from the literature (MIRI and NeMars) in finding the peak electron density value, and the peak density height in terms of root-mean-square error and mean absolute error.
\end{abstract}


\section{Introduction}
The Martian ionosphere is an active layer composed of ions and electrons produced by solar radiation. This layer has been widely explored through radio science experiments onboard several Martian missions, leading to a better understanding of its dynamic nature. A remarkable investigation of the Martian ionosphere began in December 1998 when the radio occultation experiment on-board Mars Global Surveyor (MGS) \cite{albee2001overview} measured 5600 electron density (ED) profiles from December 1998 to June 2005, covering a major portion of the 23rd solar cycle. The ED profiles retrieved from MGS have been used diversely by several researchers to study the Martian ionosphere, resulting in many studies in the literature \cite{sanchez2013nemars,mendillo2013new}.\par

The ED profile provides crucial information about the main features of the Martian ionosphere such as peak ED, peak density height, and the total electron content. Therefore, developing models to predict ED is critical for ionospheric studies. This work aims to utilize a novel approach by employing a Machine Learning (ML) model to predict the ED of the Martian ionosphere using MGS observations. The constructed model should predict the ED by solving a regression problem emerging from the predictors within the dataset. The success of this approach will contribute to opening new horizons for ML enabled ionospheric studies on Mars.\par

\subsection{Literature Review}
Three Martian ionosphere models will be discussed and compared with the proposed model in this work as they are the most relevant. The first is the \emph{NeMars} model, which was presented in \cite{sanchez2013nemars} as an empirical model for the dayside ED of the Martian ionosphere. It relied mainly on active ionospheric sounding from the MEX mission, and to a lesser extent, radio occultation data from the MGS mission. The NeMars model derived an empirical expression for the ED and peak altitude of the main and secondary ionization peaks by considering the solar zenith angle (SZA), solar flux ($F_{10.7}$), and heliocentric distance ($d$).\par

The second model is the Mars Initial Reference Ionosphere (MIRI), which is a semi-empirical model first proposed in \cite{mendillo2013new} (\emph{MIRI-2013}). Unlike NeMars, MIRI-2013 is only based on data from the MEX mission and not MGS. However, similar to NeMars, it derived an empirical expression for the ED of the main ionization peak by considering SZA, $F_{10.7}$, and $d$. This empirical expression was updated in \cite{mendillo2018mars} (\emph{MIRI-2018}), and was validated with MAVEN radio occultation data.\par


Several previous works have noted that the Martian ionosphere can be represented to an extent by Chapman-type layers \cite{mendillo2011modeling,nvemec2011dayside}. However, as the Chapman grazing incidence approximation is invalid for SZA exceeding $75 \degree$ \cite{smith1972numerical}, other representations are required for SZA ranging from $70-90\degree$. The model proposed in this work targets this region of the ionosphere. We use a ML approach to train and test a model that is purely based on MGS data, and compare it to the MIRI and NeMars models. The decision to rely purely on MGS data for this work stems from two main reasons. First, MGS was the only mission to cover most of the 23rd solar cycle, which was more severe than the 24th solar cycle. Second, the purpose of this work is to create a model for high SZA conditions, and since SZA of the MGS observations ranges from $70-90\degree$, using it as the source of this work was only natural.\par

\section{Dataset}
This work is based on data collected by the MGS mission \cite{albee2001overview}. The variables used include the solar zenith angle, solar longitude, local true solar time, altitude, latitude, longitude, and ED (see Table \ref{table:1}). Each predictor was chosen due to its importance in the prediction process. The altitude, longitude, and latitude variables help in defining spatial variations. Furthermore, the local true solar time and solar longitude define the temporal variations caused by diurnal and seasonal changes, respectively. The SZA, a commonly used predictor of ED, was chosen as it represents the angle at which solar radiation reaches the upper atmosphere \cite{sanchez2013nemars}. Additionally, a measure of solar activity was added in the form of the daily total sunspot number (SSN), retrieved from \cite{sidc}. This work aims to predict the ED using the seven other variables as predictors.\par
In total, the number of ED profiles provided by MGS was $5600$. Out of these, $220$ profiles of the southern hemisphere were discarded, as they represented less than $5\%$ of the total dataset. The remaining $5380$ profiles were randomly split for training and testing using the holdout technique, where the ratio of training to testing data was set to $9:1$. Furthermore, the training data was split for validation using $9$-fold cross-validation. All negative values of ED were discarded from the training and testing datasets. Finally, we should mention that the training and testing were done using MATLAB R2021a on a machine with an Intel Xeon (8-core) processor and 32 GB of RAM.\par

\begin{table}[t!]
\caption{List and statistics of variables used}
\begin{center}
\begin{tabular}{|l|c|c|c|}
\hline
\textbf{Variable} & \textbf{Max.} & \textbf{Mean} & \textbf{Min.}\\
\hline
SSN & 352 & 94.4 & 0 \\ \hline
Solar Longitude (Degree) & 227.3 & 149.2 & 70.2 \\ \hline
Solar Zenith Angle (Degree) & 89.2 & 76.7 & 71.0 \\ \hline
Local Solar Time (Hour) & 14.7 & 8.6 & 2.8 \\ \hline
Altitude (km) & 246.2 & 155.9 & 71.6 \\ \hline
Latitude of Profile (Degree) & 85.5 & 73.6 & 60.6 \\ \hline
Longitude of Profile (Degree) & 360 & 178.4 & 0.1 \\ \hline
ED ($\times 10^9 / m^3$) & 132.3 & 37.4 & 0.8 \\ 
\hline
\end{tabular}
\label{table:1}
\end{center}
\end{table}

\section{Martian Ionosphere Models}

\subsection{NeMars}
The \emph{NeMars} empirical model was proposed in \cite{sanchez2013nemars}. The empirical expression for the maximum ED $N_{\text{max}}$ of the main ionization peak in electrons per $m^3$ is given as
\begin{equation}
\label{eqn:1}
N_{\text{max}}^\text{NeMars}=\frac{10^8}{d}\left(4.5\,F_{10.7}^{E}-190\exp{\left(\frac{\text{SZA}}{37.2}\right)} + 2300\right),
\end{equation}
where $d$ is the heliocentric distance in AU, and $F_{10.7}^{E}$ is the solar flux index, as observed on Earth, and is available at \url{https://omniweb.gsfc.nasa.gov/form/dx1.html}. The main peak height can be expressed as
\begin{equation}
\label{eqn:2}
h_{\text{peak}}^\text{NeMars}=0.001 \exp{\left(\frac{\text{SZA}}{8.3}\right)}+132.4,
\end{equation}
in kilometers.

\subsection{MIRI}
The Mars Initial Reference Ionosphere (MIRI) is a semi-empirical model first proposed in \cite{mendillo2013new} (\emph{MIRI-2013}), and updated in \cite{mendillo2018mars} (\emph{MIRI-2018}). The maximum ED for these two models is given by
\begin{equation}
\label{eqn:3}
N_{\text{max}}^\text{MIRI-2013}=\frac{1.524}{d}\left(2.58 \times 10^{10} \sqrt{F_{10.7}^{Me} \times \cos{\left(\text{SZA}\right)}}\right),
\end{equation}

\begin{equation}
\label{eqn:4}
\resizebox{.95\hsize}{!}{$
N_{\text{max}}^\text{MIRI-2018}=\frac{1.524}{d}\left(2.23 \times 10^{10} \sqrt{F_{10.7}^{Me} \times \cos{\left(\text{SZA}\right)}} + 1.28\right),$}
\end{equation}
and represented by electrons per $m^3$. The $F_{10.7}^{Me}$ is the effective solar flux and is characterized by
\begin{equation}
\label{eqn:5}
F_{10.7}^{Me}=\frac{F_{10.7}^{Md}+F_{10.7}^{Ma}}{2d^2},
\end{equation}
where $F_{10.7}^{Md}$ corresponds to the $F_{10.7}^{E}$ value when the side of the Sun facing Mars was observed on Earth \cite{mendillo2018mars} (with a correction of $\pm14$ days), and $F_{10.7}^{Ma}$ is the three solar rotation, i.e., $81$-day average.\par 
As the MIRI models are not as straightforward as NeMars, to ensure the correct reproduction of the MIRI models, the input parameters and the output $N_{\text{max}}^\text{MIRI}$ of our reproduction was confirmed with the online version of the MIRI model, available at  \url{http://sirius.bu.edu/miri/miri.php}.\par

\subsection{Proposed Electron Density Model}
We evaluated some of the more widely used ML techniques in terms of root-mean-square error (RMSE), mean absolute error (MAE), and coefficient of determination (R$^2$),
\begin{equation}\label{eq1}
\text{RMSE} = \sqrt{\frac{\sum_{i=1}^{N}(x_i - y_i)^2}{N}},
\end{equation}
\begin{equation}\label{eq2}
\text{MAE} = \frac{\sum_{i=1}^{N} |x_i - y_i| }{N},
\end{equation}
\begin{equation}
\text{R}^2 = 1 - \frac{\sum\limits_{i=1}^N(x_i-y_i)}{\sum\limits_{i=1}^N(x_i-\bar{x})^2},
\label{eq3}
\end{equation}
where $x_i$ is the observed (true) value of ED, $y_i$ is the predicted value of ED, $N$ is the total number of samples, and $\bar{x}$ is the mean of the samples.\par

\begin{table}[t!]
\caption{Performance of the evaluated machine learning methods (best values in each category are in \textbf{bold})}
\begin{center}
\adjustbox{max width=\columnwidth}{
\begin{tabular}{|l|l|c|c|c|}
\hline
\textbf{Method} & \textbf{Parameters} & \textbf{RMSE} & \textbf{R}$^2$ & \textbf{MAE} \\
\hline
LR & - & $23.327$ & $0.18$ & $18.696$ \\ \hline
\multirow{3}{*}{RT} & min. leaf size = $4$ & $4.328$ & $0.97$ & $3.178$ \\ \cline{2-5} 
 & min. leaf size = $12$ & $4.554$ & $0.97$ & $3.359$ \\ \cline{2-5} 
 & min. leaf size = $36$ & $5.000$ & $0.96$ & $3.718$ \\ \hline
\multirow{6}{*}{SVR} & kernel = linear & $24.016$ & $0.13$ & $17.985$ \\ \cline{2-5}
 & kernel = quadratic & $16.029$ & $0.61$ & $11.838$ \\ \cline{2-5}
 & kernel = cubic & $11.099$ & $0.81$ & $8.017$ \\ \cline{2-5}
 & kernel = gaus., scale = $0.61$ & $4.136$ & $0.97$ & $3.036$ \\ \cline{2-5}
 & kernel = gaus., scale = $2.4$ & $7.528$ & $0.91$ & $5.536$ \\ \cline{2-5}
 & kernel = gaus., scale = $9.8$ & $12.745$ & $0.75$ & $9.198$ \\ \hline
Boosted RT & min. leaf size = 8 & $7.330$ & $0.92$ & $5.450$ \\ \hline
Bagged RT & min. leaf size = 8 & $\mathbf{3.498}$ & $\mathbf{0.98}$ & $\mathbf{2.602}$ \\ \hline
GPR & kernel = squared exp. & $6.050$ & $0.94$ & $4.452$ \\
\hline
\end{tabular}}
\label{table:2}
\end{center}
\end{table}

The evaluated methods were: Linear Regression (LR), Regression Tree (RT), Support Vector Regression (SVR), Boosted RT, Bagged RT (BRT), and Gaussian Process Regression (GPR). We will be introducing the RT and BRT methods in the following paragraphs. For more information regarding the other utilized ML techniques, the reader is referred to \cite{friedman2001elements}.\par

A RT consists of nodes, branches, and leaves. It is constructed by splitting data iteratively into branches to form a hierarchical structure. Each node splits into only two branches, and each branch connects between two nodes. The upper edge of the tree starts from the root node, and the lower edge of the tree consists of leaf nodes. By following a path from the root node to the leaf nodes, the user can obtain a prediction value from the RT method \cite{sutton2005classification}.\par 

Bagging, or bootstrap aggregation, was first proposed by \cite{breiman1996bagging}. It is a method where multiple versions of a predictor, in this case a RT, are generated with each of them generating their own predictions, and the final prediction is the average of all of these individual predictions.\par

The performance of the evaluated ML methods is included in Table \ref{table:2}. As is clear from Table \ref{table:2}, the BRT was the best performing method in terms of RMSE, R$^2$, and MAE. Therefore, it was selected as the target method for this work. Next, Bayesian optimization was used to search for the optimal set of hyperparameters that minimized the RMSE. Two different sets of parameters were found (Table \ref{table:3}). The first set, minimum leaf size $= 1$ and number of learners $= 500$ achieved the best RMSE, MAE, and R$^2$ performance. However, due to the large number of learners, this resulted in a relatively slow model. The second-best set of parameters were minimum leaf size $= 1$ and number of learners $= 10$. This resulted in a much more reasonable model with a prediction speed that is at least 42 times faster, with almost similar prediction accuracy. Therefore, this optimized BRT (OBRT) method was chosen as the final model of this work\footnote{The OBRT model is available from the corresponding author, A. Darya, upon reasonable request.}. If prediction accuracy was the first priority of the reader, then we recommend using BRT with the hyperparameters of set 1.\par

\section{Results and Discussion}

\begin{table}[t!]
\caption{Performance of the two best hyperparameter sets}
\begin{center}
\begin{tabular}{|l|c|c|}
\hline
\textbf{Parameter} & \textbf{Set 1} & \textbf{Set 2} \\
\hline
Min. Leaf Size & $1$ & $1$\\ \hline
Num. of Learners & $500$ & $10$\\ \hline
RMSE & $3.156$ & $3.360$\\ \hline
R$^2$ & $0.98$ & $0.98$\\ \hline
MAE & $2.412$ & $2.538$\\ \hline
Prediction Speed (obs/sec) & $\approx{4\,000}$ & $\approx{170\,000}$\\\hline
Training Time (hours) & $\approx{4.4}$ & $\approx{0.1}$\\
\hline
\end{tabular}
\label{table:3}
\end{center}
\end{table}

\begin{figure}[htbp]
\centerline{\includegraphics[width=0.9\columnwidth]{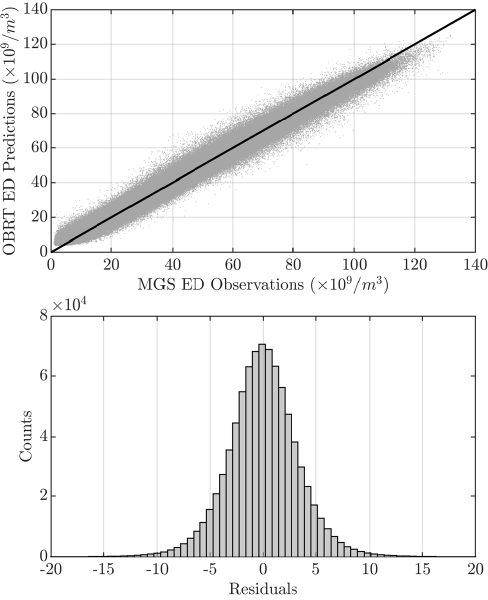}}
\caption{True response versus OBRT predicted response using testing data.}
\label{fig:1}
\end{figure}

In the top panel of Fig. \ref{fig:1}, we see the OBRT predicted response plotted against the MGS true response using the testing data. We note the proximity of the points to the perfect prediction diagonal line, indicating the accuracy of the proposed OBRT model. Furthermore, the histogram in Fig. \ref{fig:1} (bottom panel) illustrates the distribution of the residuals, i.e., the prediction errors of OBRT, where $\text{Residuals}_i=x_i-y_i$. Using data from the histogram in Fig. \ref{fig:1}, it was noted that $88\%$ of the residuals were between $-5$ and $5$ ($\times 10^9 / m^3$), and $50\%$ between $-2$ and $2$.\par

\begin{figure}[t!]
\centerline{\includegraphics[width=\columnwidth]{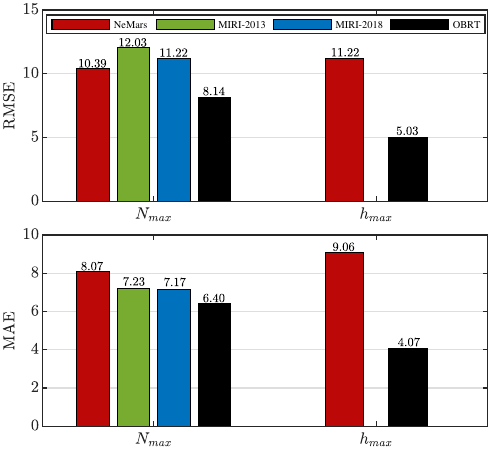}}
\caption{Comparison of the MAE and RMSE of the $h_{\text{max}}$ and $N_{\text{max}}$ model predictions.}
\label{fig:2}
\end{figure}

In Fig. \ref{fig:2}, the M2 peak ED $N_{\text{max}}$ predictions made by the NeMars, MIRI-2013, MIRI-2018, and the OBRT models are compared in terms of MAE, and RMSE. Furthermore, the $h_{\text{max}}$ predictions made by the NeMars, and OBRT models are also compared. However, as the MIRI models do not predict the $h_{\text{max}}$ values, they were not included in the second comparison.\par
As is clear from Fig. \ref{fig:2}, the proposed OBRT model outperforms all other models in both MAE and RMSE for the $N_{\text{max}}$ and the $h_{\text{max}}$ predictions, where it had less than half the MAE and RMSE of NeMars. For the $N_{\text{max}}$ predictions, both MIRI models outperform NeMars in terms of RMSE but not MAE. Furthermore, the MIRI-2018 model outperformed the older MIRI-2013 model.\par

\begin{figure}[t!]
\centerline{\includegraphics[width=\columnwidth]{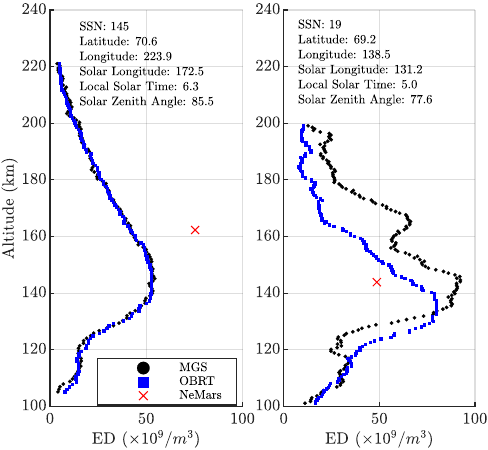}}
\caption{Left Panel: The OBRT model predicted ED profile with the lowest RMSE. Right Panel: The OBRT model predicted ED profile with the highest RMSE. The red cross represents the NeMars prediction of the $N_{\text{max}}^\text{NeMars}$ and $h_{\text{peak}}^\text{NeMars}$. In addition to altitude, the six other input parameters are shown with each case, as retrieved from MGS, and used as inputs in the OBRT model.}
\label{fig:3}
\end{figure}

In Fig. \ref{fig:3}, we present two predictions of the complete ED profile made by the proposed OBRT model. The left panel shows the prediction with the lowest RMSE, i.e., the most accurate prediction, while the right panel shows the highest RMSE prediction, i.e., the least accurate prediction. Additionally, since NeMars predicts the peak height and the peak density, it was included in the figure as a comparison. We note that in both cases, the least and most accurate predictions, OBRT outperforms NeMars in terms of $N_{\text{max}}$. However, the peak height prediction obtained by NeMars was more accurate for the highest RMSE case.\par
It is important to note that while NeMars and the MIRI models first derive the $N_{\text{max}}$ value, then fit the ED curve into it \cite{nvemec2011dayside}, the proposed OBRT model predicts every point independently, and as such, erroneous predictions of the $N_{\text{max}}$ do not necessarily mean incorrect predictions of the other segments of the ED profile.\par

\section{Conclusion}
In this work, we constructed a model of the Martian ionosphere at high SZA conditions, using data retrieved from MGS and machine learning techniques. The model was able to predict the ED profile using seven predictors which are Sunspot Number, Solar Longitude, Solar Zenith Angle, Local Solar Time, Altitude, Latitude, and Longitude. We have evaluated several machine learning methods such as Linear Regression, Regression Tree, Support Vector Regression, Boosted Regression Trees, Bagged Regression Trees (BRT), and Gaussian Process Regression. Among the mentioned models, the BRT yielded the best RMSE, R$^2$, and MAE. Furthermore, using Bayesian optimization, the optimal hyperparameters of the BRT model were found, allowing further optimization of the model. The optimized BRT model outperformed other Martian ionosphere models from the literature (MIRI and NeMars) in finding the peak ED value and the peak density height in terms of RMSE and MAE.\par

\section*{Acknowledgment}
We thank the authors of \cite{sanchez2013nemars, mendillo2013new,mendillo2018mars}, and especially Beatriz S{\'a}nchez-Cano and Clara Narvaez for assisting us in using their respective models. 

\bibliographystyle{IEEEtran.bst}
\bibliography{MGS_IEEE.bib}

\end{document}